\documentclass[conference]{IEEEtran}
\IEEEoverridecommandlockouts
\usepackage{cite}
\usepackage{amsmath,amssymb,amsfonts}
\usepackage{algorithmic}
\usepackage{graphicx}
\usepackage{textcomp}
\usepackage{xcolor}

\usepackage{booktabs}
\usepackage{multirow}
\usepackage{makecell}
\newcommand{\K}[1]{\textcolor{green!60!black}{#1}}
\newcommand{\Plaus}[1]{\textcolor{blue!70!black}{#1}}
\newcommand{\U}[1]{\textcolor{orange!90!black}{#1}}
\newcommand{\NP}[1]{\textcolor{red!70!black}{#1}}

\usepackage{fancyhdr} 
\fancypagestyle{firstpage}{
    \fancyhf{}
    \fancyfoot[L]{\footnotesize 979-8-3195-1905-4/26/\$31.00 \copyright 2026 IEEE. Personal use of this material is permitted. Permission from IEEE must be obtained for all other uses, in any current or future media, including reprinting/republishing this material for advertising or promotional purposes, creating new collective works, for resale or redistribution to servers or lists, or reuse of any copyrighted component of this work in other works.}

}

\def\BibTeX{{\rm B\kern-.05em{\sc i\kern-.025em b}\kern-.08em
    T\kern-.1667em\lower.7ex\hbox{E}\kern-.125emX}}
\begin{document}

\title{Information-Theoretic Causal Modelling of Semiconductor Process Dynamics}

\author{\IEEEauthorblockN{Daniel Sørensen\IEEEauthorrefmark{1}\IEEEauthorrefmark{2}, Giorgio Melchiorre\IEEEauthorrefmark{2}, Sudip Bandyopadhyay\IEEEauthorrefmark{2}, \\Sandip Halder\IEEEauthorrefmark{2}, Roel Wuyts\IEEEauthorrefmark{1}\IEEEauthorrefmark{2}, and Bappaditya Dey\IEEEauthorrefmark{2}}
\IEEEauthorblockA{\IEEEauthorrefmark{1}Dept. Computer Science, KU Leuven, 3000 Leuven, Belgium}
\IEEEauthorblockA{\IEEEauthorrefmark{2}imec, Kapeldreef 75, 3001 Leuven, Belgium\\
Email: \{daniel.sorensen, giorgio.melchiorre, sudip.bandyopadhyay, sandip.halder, bappaditya.dey, roel.wuyts\}@imec.be}
}


\maketitle

\thispagestyle{firstpage}

\begin{abstract}
With the progress of the semiconductor industry toward increasingly complex compute devices and tighter process tolerances, advanced process control has become crucial. This work explores a novel framework to infer the underlying dynamics of semiconductor processes, directly from raw equipment log-file time-series data. By modelling the tool dynamics as a stochastic dynamical system comprising (a) a deterministic component and (b) a stochastic component, we estimate entropy transfer rates between variables through the Liang-Kleeman and Pires formalism. Preliminary results indicated that 7.5\% of the inferred dependencies were known, 36.0\% were plausible, 17.5\% represented previously uncharacterised relationships, and 39.0\% were inconsistent with established process knowledge. These findings demonstrate the framework’s capability to uncover novel causal insights, while motivating further improvements to reduce inconsistent findings.
\end{abstract}

\begin{IEEEkeywords}
Causal Inference, Information Theory, Multivariate Time Series, Semiconductor Manufacturing, Process Dynamics
\end{IEEEkeywords}

\section{Introduction}
In the pursuit of ever more miniaturised computing devices \cite{roadmap}, semiconductor manufacturing has evolved toward increasingly complex process flows with tighter tolerances, making advanced process control crucial for high volume production. Consequently, Fault Detection and Classification (FDC) systems deploy sensor arrays measuring key parameters such as chamber temperature and gas-flow rate, producing multivariate time-series (MTS) data for analysis. Despite their importance, current FDC approaches face several limitations: \textbf{1.} reliance on statistical thresholds, rule‑based alarms, and manual expert investigation, limiting the generalisability across tools or chambers; \textbf{2.} limited dynamic process modelling capabilities such as time‑dependency and non-linear relationships, instead focusing on static summary statistics; \textbf{3.} struggle to represent high‑dimensional interactions, frequently treating correlated signals as independent or relying on simplified correlation models; \textbf{4.} dependency on supervised learning, requiring "labelled fault data" resulting in failed detection of previously unseen anomalies, and limited sensitivity to gradual process drifts, equipment ageing, or recipe modifications that can produce false alarms. These delayed corrections cause unplanned tool downtime costing up to \$100K per hour \cite{DowntimeCosts}. This motivates modelling frameworks that capture complex, high-dimensional, dynamic interactions among process variables while enabling interpretable root-cause analysis.

 This work proposes a novel information-theoretic framework for modelling semiconductor manufacturing processes as stochastic dynamical systems, whose governing equations are estimated directly from observed multivariate time-series data. The system dynamics are decomposed into two components: (i) \texttt{deterministic}, approximated using wavelet basis functions to effectively capture both non-smooth dynamics and high-noise signals, and (ii) \texttt{stochastic}, modelled using Gaussian basis functions. Based on the estimated system dynamics, we compute the rate of entropy transfer between variables to quantify causal dependencies, enabling the construction of interpretable causal graphs characterising underlying process interactions. This fully unsupervised framework, requiring no prior process knowledge, provides novel insights into process dependencies, distinguishes between deterministic and stochastic interactions, and establishes a foundation for counterfactual analyses and data-driven process optimisation.
 
 Section \ref{sec:methods} describes the formalism, Section \ref{sec:R&D} presents the results, Section \ref{sec:limit} discusses them and the study's limitations, and Section \ref{sec:conclusion} concludes with future work.

 \section{Methods}
\label{sec:methods}

\subsection{Information Theory foundation}
Let us consider a $D-$dimensional dynamical system described by the state vector $\textbf{X}=\left(X_1, ..., X_D\right)$ and its associated probability density function (pdf) $\rho_{\textbf{X}}$, whose evolution is governed by the stochastic differential equation in \eqref{eq:sde}:
\begin{equation}
    dX_i = F_i\left(\textbf{X},\boldsymbol{\theta},t\right) dt + \sum_{k=1}^D B_{i,k}\left(\textbf{X},\boldsymbol{\theta},t\right)dW_k
    \label{eq:sde}
\end{equation}
where $F_i\left(\textbf{X},\boldsymbol{\theta},t\right)$ are the deterministic components of the $D-$dimensional vector field $\textbf{F}$, representing the underlying mechanics driving the system, $\boldsymbol{\theta}$ denotes the parameter vector, $dW_k$ are independent Gaussian Wiener processes and $B_{i,k}$ noise-diffusion coefficients that induce stochastic forcings into the system. The noise-diffusion coefficients are collected into a differentiable matrix $\textbf{B}$, from which the positive-definite matrix of probability density function diffusivities is constructed as $\textbf{G}\equiv\textbf{BB}^T$, with elements $g_{i,j} \equiv \sum_{k=1}^D B_{i,k}B_{j,k}$.

The formalism developed by Liang and Kleeman \cite{Liang2016,Liang2005} address such systems through the concept of Shannon entropy, $H_{X_i} = E[-log \rho_i]$ associated with the variable $X_i$, and its rate of entropy change (REC) $\frac{dH_{X_i}}{dt}$, where $\rho_i$ is the pdf of $X_i$ and E[.] is the expectation operator. A positive (negative) rate indicates a loss (gain) of information. The authors demonstrated that $\frac{dH_{X_i}}{dt}$ can be decomposed into a global rate of entropy transfer (RET), from the complementary vector $X_{\sim i} = \textbf{X}/\{X_i\}$, and self-entropy generation (SEG) as expressed in \eqref{eq:REC_decomp}. 
\begin{equation}
    \frac{dH_{X_i}}{dt} = RET+SEG=T\left(X_{\sim i}\rightarrow X_i\right) + \frac{dH_{X_i,self}}{dt}
    \label{eq:REC_decomp}
\end{equation}
 Pires \textit{et al.} \cite{Pires2024} extend this formalism by decomposing the global RET into a sum of direct contributions from individual variables and a synergetic term arising from non-null conditional covariances due to non-linearities in $F_i$ and $g_{i,j}$.
 \begin{equation}
    T(X_{\sim i} \rightarrow X_i) = \sum_{j \neq i}T(X_j \rightarrow X_i) + T(X_{\sim i} \rightarrow X_i)_{syn}
    \label{eq:global-single-RET}
\end{equation}

Furthermore, if the functions $F_i$ and $g_{i,i}$ can be expressed as a product-separable function, according to \eqref{eq:prod-sep}. 
\begin{equation}
    f(\textbf{X},t) =\sum_l \theta_{l,i}\prod_r \Psi_{l,r}(X_r,t)\quad \text{for }f \in \{F_i,\,g_{i,i}\}
    \label{eq:prod-sep}
\end{equation}

\begin{table}[b]
\renewcommand{\arraystretch}{1.3}
\centering
\caption{Median train and test MSE across all variables for each dataset.}

\begin{tabular}{
    c
    @{\hspace{15pt}}c
    @{\hspace{15pt}}c
    @{\hspace{19pt}}c
    @{\hspace{19pt}}c
}
\toprule
\textbf{Tool} &
\textbf{Side} &
$\textbf{Recipe}_A$ &
$\textbf{Recipe}_B$ &
$\textbf{Recipe}_C$ \\
& & \small (Train\,\textbar{}\,Test) & \small (Train\,\textbar{}\,Test) & \small (Train\,\textbar{}\,Test) \\
\midrule

\multirow{2}{*}{Tool$_1$}
 & 1
   & 0.009 \textbar{} 0.010
   & 0.117 \textbar{} 0.117
   & \textbf{0.197 \textbar{} 0.547} \\
 & 2
   & \textbf{0.221 \textbar{} 0.236}
   & 0.042 \textbar{} 0.050
   & \textbf{0.185 \textbar{} 0.562} \\
\midrule

\multirow{2}{*}{Tool$_2$}
 & 1
   & 0.046 \textbar{} 0.050
   & 0.021 \textbar{} 0.022
   & 0.009 \textbar{} 0.009 \\
 & 2
   & 0.008 \textbar{} 0.008
   & \textbf{0.225 \textbar{} 0.216}
   & 0.028 \textbar{} 0.025 \\
\bottomrule

\end{tabular}
\label{tab:median_fit_results}
\end{table}

then the RET components can be split into a deterministic term and a stochastic term as follows:
\begin{subequations}
    
    \begin{equation}
        \begin{split}
            &T(X_{j} \rightarrow X_i)_F =\\
            &= \sum_l \theta_{l,i} E \left[\Psi_{l,i}\frac{d E(\Psi_{l,j}|X_i)}{dX_i} E\left(\prod_{r}\Psi_{l,r}|X_i\right)\right]
        \end{split}
    \label{eq:s-RETf-sep}
    \end{equation}
    \\
    \begin{equation}
        \begin{split}
            &T(X_{j} \rightarrow X_i)_g = \\
            &=\sum_l \frac{\varphi_{l,i}}{2} E \left[\Phi_{l,i}\frac{d^2 E(\Phi_{l,j}|X_i)}{dX_i^2} E\left(\prod_{r}\Phi_{l,r}|X_i\right)\right]
        \end{split}
    \label{eq:s-RETg-sep}
    \end{equation}
    
    \label{eq:RETs-SEGs}
\end{subequations}
where the $i$-index denotes the consequential variable or a function dependent on it, the $j$-index denotes the causal variable, and $r\equiv \sim(i,j)$ denotes the contextual variables, i.e., components independent of both $i$ and $j$. These RET components provide the direct influence of one variable $(X_j)$ on another $(X_i)$, additionally, these not only have a clear physical interpretation (they measure the rate of information transfer in nats/unit time), but also have the property of non-bi-directionality, providing a causal interpretation, contrary to metrics such as the Pearson correlation coefficient. For the remaining entropy rate description of the system we recommend reading the work by Pires et al. \cite{Pires2024}.

\subsection{RET numerical computation}
\label{numerical}
 Since the complete driving equations of semiconductor process tools are generally unknown, these must be estimated from the corresponding multivariate time series data recorded from the process tools' log files. We estimate $\hat{F}_i(\textbf{X})$ by regression of the numerical gradient of $X_i$, $\frac{dX_i}{dt}$, while $\hat{g}_{i,i}(\textbf{X})$ is subsequently estimated by regression of the obtained squared residuals: $\varepsilon^2_i \equiv \left(\frac{dX_i}{dt}-\hat{F}_i\right)^2$. Due to the presence of non-smooth signals in semiconductor processing, we utilise Ricker wavelets \eqref{eq:ricker} for fitting $F_i$, permitting a flexible regression of both smooth and non-smooth derivatives. By exploiting the almost compact support of the wavelets, we first efficiently construct a wavelet frame covering the data, followed by a Stepwise Selection by Orthogonalisation (SSO) procedure to select only the $S=30$ most relevant wavelet candidates. Lastly, we fine tune the wavelet parameters of the selected candidates by training a Wavelet Neural Network (WNN). We follow the implementations given in \cite{QinghuaZhang1997, Alexandridis2013}. 
\begin{equation}
    \begin{split}
       &\psi_{m,n}(x)= \left(1-z_{m,n}^2 \right)e^{-\frac{z_{m,n}^2}{2}},\\
       & z_{m,n}=2^m*x-2n, \quad m,n \in \mathbb{Z}
        \label{eq:ricker} 
    \end{split}
\end{equation}

 Albeit its advantages, wavelet based regression faces two numerical challenges: (i) the number of possible wavelet combinations for constructing the product in \eqref{eq:prod-sep} grows with $(MN)^D$, with $M,N$ the number of scale and translation indices, and (ii) for high-dimensional systems ($D>6$), the product terms in \eqref{eq:prod-sep} quickly tend toward 0, prohibiting any function estimation. Therefore, we restrict the products in \eqref{eq:prod-sep} to a maximum interaction degree $deg=4$, and restrict the scale index to $m\in \{3,4,5,6\}$ ($M=4$), given the data is normalised to $[0,1]$.
 
The $\hat{g}_{i,i}$ are fitted following the algorithm described in Appendix B.1 of \cite{Pires2024}, using Gaussian basis functions up to degree 2 \eqref{eq:prod-sep}.
 
 Once $F_i$ and $g_{i,i}$ have been decomposed into basis functions, the RET terms in \eqref{eq:RETs-SEGs} can be computed very efficiently. These terms consist of expectations, which are evaluated as averages over the observed data, and conditional expectations, estimated via local polynomial regression \cite{Fan2018} (implemented in \cite{kernreg}), with the addition of linear interpolation for highly sparse regions to prevent gradient overestimation.

Finally, to assess the significance of the obtained RET estimations, we follow the methodology of \cite{Pinto2024}, and perform surrogate testing. For each signal $X_i$, we generate 100 surrogate time-series using the IAAFT method \cite{Venema2006} and recompute $T(X_j \rightarrow X_i)$ using surrogate realisations of $X_j$. If the true RET yields a $p\text{-value}<\alpha$, with $\alpha=0.05$, it is considered statistically significant.

\subsection{Data description}
\label{sec:data}
To evaluate the proposed methodology, log-file data was acquired from three process recipes run on two matched production (FAB) tools, each with two simultaneously operating chamber sides. Due to confidentiality constraints, details regarding tool model, tool types, recipes and variable names are anonymised as $Tool_1/Tool_2$, $Recipe_A/Recipe_B/Recipe_C$ and $Var\_i$, where $i$ denotes an integer index. Although the two tools are intended to be matched, process engineers have reported minor discrepancies in wafer properties measured in subsequent process steps. Consequently, the collected data were partitioned according to \{tool:2, side:2, recipe:3\}, resulting in a total of 12 datasets. Each dataset contains between 50 and 108 MTS instances, $Tool_1$ datasets contained 34 active variables and $Tool_2$ contained 42 variables. The average running lengths for recipes $A$, $B$ and $C$ were 86, 47 and 51 seconds, respectively.  In total, 739 MTS instances were collected across the different tool/side/recipe combinations.

\section{Results}
\label{sec:R&D}

\subsection{Wavelet Regression}
\label{subsec:result-wavelet}
The wavelet regression was performed on 90/10 train/test split, as commonly used in related literature. To assess the quality of the regression fits, we report the median of mean squared errors (MSE) for each dataset. Concretely, we first compute the MSE for each variable across all instances, followed by the median MSE across the variables. From Table \ref{tab:median_fit_results}, it is observed that ten of the datasets exhibited comparable train and test errors, indicating stable generalisation. Additionally, for seven of these ten datasets exceptionally low errors $\le0.05$ were obtained. However, for $Tool_1\text{-}Recipe_C$, the test error was $> 2.5\times$ larger than the training error, while errors exceeding $0.2$ were obtained for two other datasets. 

\begin{figure}[t]
    \centering
    \includegraphics[width=0.49\textwidth]{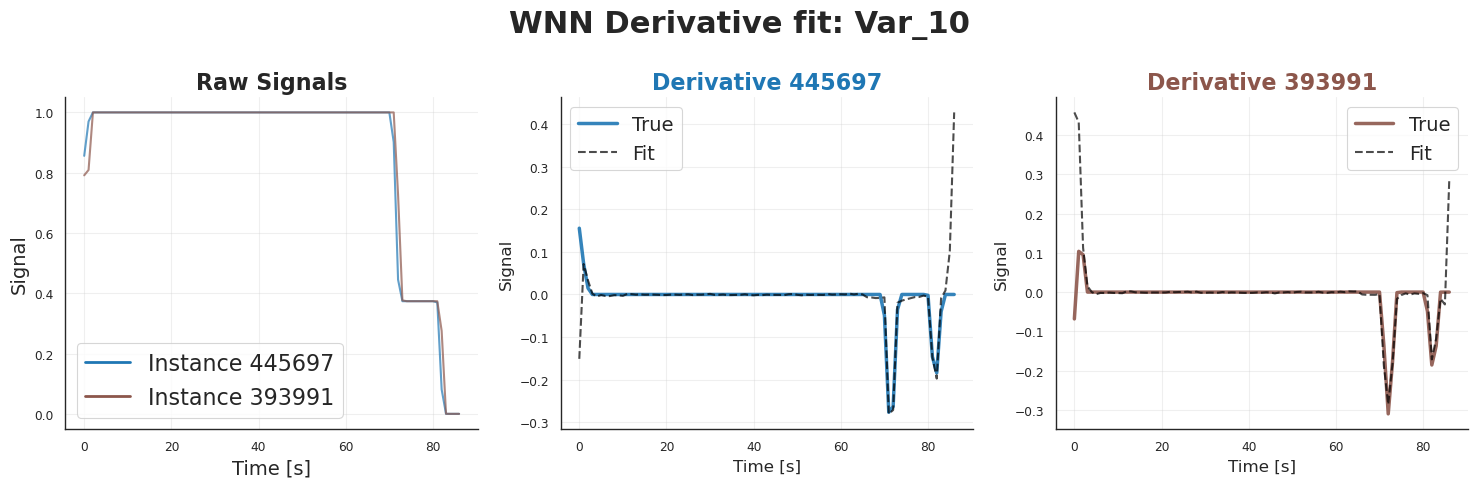}
    \includegraphics[width=0.49\textwidth]{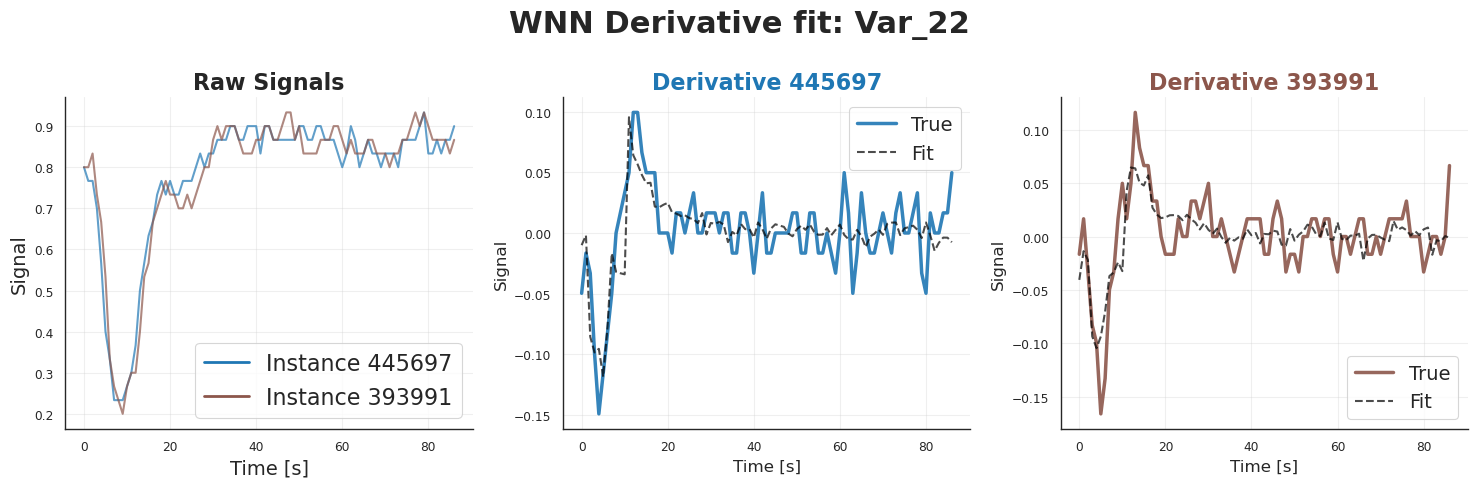}

    \caption{
        Estimation of $F_i$ using wavelet neural network regression of the derivatives of two variables: $Var\_10$, a low noise, non-smooth signal, and $Var\_22$, a smooth, but noisy signal. Both signals are drawn from the $Tool_1/Side_1/Recipe_A$ dataset}
    \label{fig:wnnfits}
\end{figure}

Two representative wavelet regression results from the dataset $Tool_1/Side_1/Recipe_A$ are presented in Fig. \ref{fig:wnnfits}. The results demonstrate the proposed methodology successfully captures both the peak-shaped derivatives associated with non-smooth transitions in variable \textit{Var\_10}, and the behaviour of the noisy signal in \textit{Var\_22}, without exhibiting over-fitting.

\subsection{RET-based Causal Graphs}
\label{subsec:result-graphs}
For each dataset, the sum of deterministic and stochastic RET values forms a square matrix which is interpreted as the adjacency matrix of a directed causal graph with RET values for weights. An average of 230 edges per graph was obtained, resulting in a total of 2742 edges across all datasets. Since causal graph generation was fully unsupervised, with no knowledge of variable semantics, we validate the inferred dependencies through a blind evaluation protocol. The 20 edges with the highest associated RET-values, from each dataset, are presented to a domain expert to evaluate them based on physical plausibility and consistency with known dependencies, based on their knowledge of semiconductor tool operation. Each edge was classified into one of four categories: \K{\textbf{Known (K)}}: representing empirically proven dependencies recognised by process engineers; \Plaus{\textbf{Plausible (P)}}: representing expected dependencies although not explicitly verified; \U{\textbf{Unknown (U)}}: representing dependencies that are unexpected but not impossible; or \NP{\textbf{Not Possible (NP)}}: representing dependencies that are considered physically impossible given the current knowledge of the tool. The validation results, summarised in Table \ref{tab:classification}, indicate the proposed methodology successfully identifies a number of meaningful process dependencies, while also revealing a substantial set of inferred relationships that are inconsistent with existing process knowledge. It should be noted that the categories \Plaus{P}, \U{U}, and \NP{NP} reflect the current understanding and experience of the evaluating process expert, rather than representing absolute physical truth. Semiconductor manufacturing systems are inherently complex, and many variable interactions remain only partially characterised within existing process knowledge. Consequently, dependencies classified as \Plaus{P}, \U{U}, or \NP{NP} may still correspond to genuine but previously unrecognised process relationships, or they represent interactions that require further experimental verification. Therefore, these classifications should be interpreted as expert-informed assessments based on current state of knowledge, and the validation outcomes may evolve as additional process insights and empirical studies become available. Investigating such potential dependencies through further experimental and analytical studies constitutes an important direction for future research.

\begin{table}[t]
\renewcommand{\arraystretch}{1.4}
\centering
\caption{Classification results on each dataset. 
Classes are \K{Known (K)}, \Plaus{Plausible (P)}, 
\U{Unknown (U)}, and \NP{Not Possible (NP)}. For each dataset the number of values per class are shown, and for the total the percentage is shown,  presented as \,\K{K} \textbar{} \Plaus{P} \textbar{} \U{U} \textbar{} \NP{NP}.}
\resizebox{\columnwidth}{!}{%
\begin{tabular}{ccccc}
\hline
\multicolumn{1}{c}{\textbf{Tool}} &
  \textbf{Side} &
  \textbf{$\textbf{Recipe}_A$} &
  \textbf{$\textbf{Recipe}_B$} &
  \textbf{$\textbf{Recipe}_C$} \\ \hline
\multirow{2}{*}{Tool$_1$} &
  1 &
  \K{2} \textbar{} \Plaus{12} \textbar{} \U{1} \textbar{} \NP{5} &
  \K{1} \textbar{} \Plaus{7} \textbar{} \U{6} \textbar{} \NP{6} &
  \K{1} \textbar{} \Plaus{3} \textbar{} \U{3} \textbar{} \NP{13} \\
 &
  2 &
  \K{2} \textbar{} \Plaus{9} \textbar{} \U{1} \textbar{} \NP{8} &
  \K{4} \textbar{} \Plaus{5} \textbar{} \U{6} \textbar{} \NP{5} &
  \K{3} \textbar{} \Plaus{9} \textbar{} \U{2} \textbar{} \NP{6} \\ \hline
\multirow{2}{*}{Tool$_2$} &
  1 &
  \K{1} \textbar{} \Plaus{6} \textbar{} \U{3} \textbar{} \NP{10} &
  \K{0} \textbar{} \Plaus{11} \textbar{} \U{5} \textbar{} \NP{4} &
  \K{3} \textbar{} \Plaus{4} \textbar{} \U{6} \textbar{} \NP{7} \\
 &
  2 &
  \K{0} \textbar{} \Plaus{10} \textbar{} \U{3} \textbar{} \NP{7} &
  \K{0} \textbar{} \Plaus{8} \textbar{} \U{2} \textbar{} \NP{10} &
  \K{1} \textbar{} \Plaus{2} \textbar{} \U{4} \textbar{} \NP{13} \\ \hline
 &
   &
   &
   &
   \\ \hline
\multicolumn{2}{c}{\multirow{2}{*}{\textbf{Total}}} &
  \multicolumn{3}{c}{\multirow{2}{*}{\K{7.5\%} \,\textbar{}\, \Plaus{36.0\%}\, \textbar{}\,\,\U{17.5\%} \,\textbar{}\, \NP{39.0\%}}} \\
\multicolumn{2}{c}{} &
  \multicolumn{3}{c}{} \\ \hline
\end{tabular}%
}
\label{tab:classification}
\end{table}

\begin{figure}[t]
    \centering
    \includegraphics[width=\columnwidth]{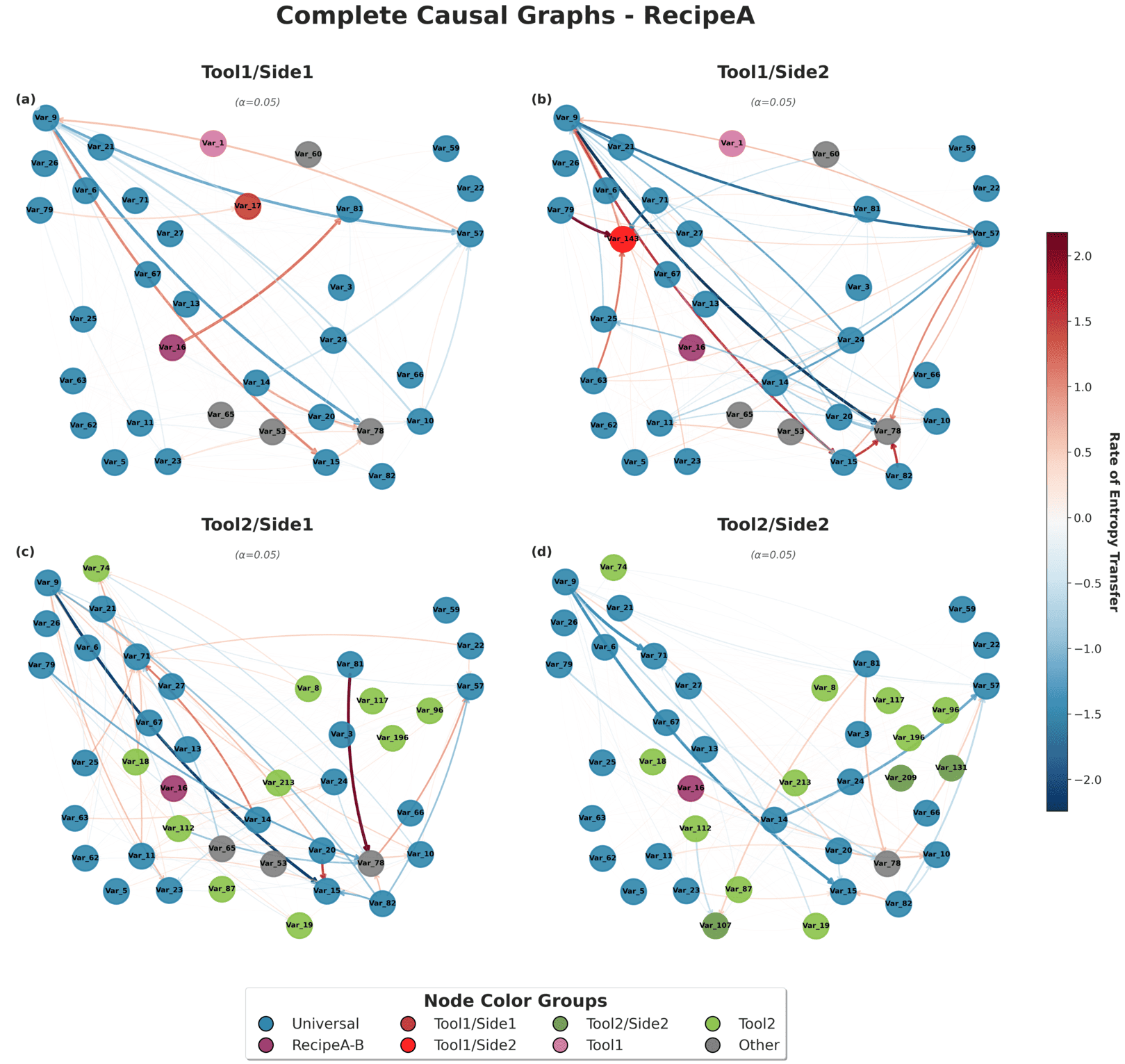}
    \caption{Comparison of the complete RET graphs from $Recipe_A$ across tool/side combinations. Nodes are colour-coded according to their presence across all datasets.}
    \label{fig:RecipeGraphs}
\end{figure}

Furthermore, since the same recipe executed across tools and sides has similar process dynamics, we expect to obtain similar graphs depicting the core dependencies of the recipe. Therefore, we analysed the agreement of graphs obtained for each recipe, and found that at most 10\% of edges appear in at least three of the graphs, while at least 64\% of edges appear in only one of the graphs. Fig. \ref{fig:RecipeGraphs} depicts this distribution for $Recipe_A$. This indicates the proposed methodology consistently did not find highly agreeing graphs for a given recipe across tools and sides.

\section{Discussion and Limitations}
\label{sec:limit}

The performance of the proposed framework is largely attributed to the ability of the wavelet-based regression to accurately fit both peak-shaped derivatives from non-smooth signals, and smooth, noise-afflicted signals. However, the wavelet regression also introduces variability and inconsistencies in the obtained results. We identified the selection procedure, SSO \cite{QinghuaZhang1997}, of $S=30$ wavelets out of $>20$ million candidates (from the generated wavelet frame), being responsible for both the variance in regression quality identified in \ref{subsec:result-wavelet} and the graph disagreement detected in \ref{subsec:result-graphs}. The large number of candidate wavelets generated during frame construction causes the selection algorithm to be sensitive to minor dataset variations. This sensitivity leads to (i) including suboptimal variable sets which are unable to reach a low error regression during the WNN training, and (ii) including varying variable sets which in turn produce graphs with few edges in common, even for similar datasets. While increasing the subset size ($S$) could potentially mitigate the variability, the SSO algorithm, despite being presented as efficient, becomes computationally expensive. Further increasing $S$ would be prohibitive for practical application in the FAB. Therefore, future research should focus on developing alternative methods to fit the derivatives with the same expressive power as wavelet basis functions while reducing the sensitivity to minor dataset variations. 

The reported proportions characterise only the most strongly weighted edges as assessed by a single expert, and the high Not-Possible fraction co-occurs with the cross-tool graph variability traced to SSO sensitivity; they should therefore be read as a feasibility indication rather than a validated measure of full-graph accuracy.

\section{Conclusion}
\label{sec:conclusion}

In this work, we approached semiconductor manufacturing process modelling from an information theory perspective to infer causal dependencies between process parameters. By leveraging the expressive power of wavelet basis functions, our proposed framework is capable of detecting both known and plausible dependencies, approximately 43\% of the largest inferred relationships, as validated by a domain expert, while identifying previously unrecognised interactions for further exploration. This study establishes a foundation for rigorous and interpretable modelling of semiconductor manufacturing dynamics, providing a systematic framework for understanding complex process dependencies and informing future process optimisation efforts. Future work will target graph stability, via stability-selection or bootstrap alternatives to SSO and false-discovery-rate control, and will benchmark against Transfer Entropy and validate on synthetic systems with known dynamics.

\bibliographystyle{IEEEtran}
\bibliography{bibfile}

\end{document}